\documentclass[12pt]{article}
\usepackage{latexsym,amsmath,amssymb,amsbsy,graphicx}
\usepackage[space]{cite} 

\usepackage[T2A]{fontenc}
\usepackage[utf8]{inputenc}

\makeatletter\def\@biblabel#1{\hfill#1.}\makeatother

\allowdisplaybreaks
\begin {document}

\noindent\begin{minipage}{\textwidth}
\begin{center}

{\Large{\bf Kink in the dilaton-axion theory with a potential \\
possessing dilaton shift symmetry
}}\\[20pt]

{\large O.\,V. Kechkin}\\[20pt]

\parbox{.96\textwidth}{\centering\small\it
Department of General Nuclear Physics,
Faculty of Physics, M.V. Lomonosov Moscow State University, 
1(2), Leninskie gory, GSP-1, Moscow 119991, Russian Federation
\\
\ E-mail: o.v.kechkin@physics.msu.ru
}\\[1cc] 

\end{center}

{\parindent5mm The sigma model with dilaton and axion is generalized by including in it a potential that is invariant under the global transformation of the dilaton shift. In the (1 + 1)-dimensional case, a soliton is constructed, which turned out to be an axion-type kink (anti-kink) with a nontrivial distribution of the dilaton field. The solution is obtained for an arbitrary value of the dilaton-axion coupling constant, and its mass is found.
 \vspace{6pt}\par}

\textit{Keywords}: sigma model with potential, dynamic symmetry, soliton.\vspace{3pt}\par

\small PACS: 05.45.Yv, 11.10.Lm, 11.30.-j.
\vspace{20pt}\par
\end{minipage}


\section*{Introduction}
\mbox{}\vspace{-\baselineskip}

Lagrangian systems with dilaton and axion appear when considering various regimes of superstring theory \cite{Pol} - \cite{StringsM}; they have non-trivial extended symmetry groups \cite{MS} - \cite{K}. The dilaton-axion sector of such systems is described by the Lagrangian
\begin{equation}\label{1}
{\cal L}_0=\frac{1}{2}\left[\left(\partial\phi\right)^2+e^{2\alpha\phi}\left(\partial\kappa\right)^2\right],
\end{equation}
where $\phi=\phi(x)$ and $\kappa=\kappa(x)$ are the dilaton and axion fields, and $\alpha$  is the dilaton-axion coupling constant (for example, $\alpha=\frac{1}{2}$ for the case of low-energy dynamics in the heterotic string theory).  We consider a $(1+d)$-dimensional system on a flat background with the Minkowski metric $g_{\mu\nu}=g^{\mu\nu}=diag(1; -1, …, -1 )$ in Cartesian orthogonal coordinates $x=\{x^{\mu}\} =\{x^0; \, x^k\}$, where  $\mu=0, \dots, d$ and $k=1, \dots, d$.  It was set $\left(\partial\phi\right)^2 = g^{\mu\nu} \partial_{\mu}\phi \partial_{\nu}\phi$ in (\ref{1}) for compactness (and similarly for  $\left(\partial\kappa\right)^2 $).

Dilaton shift transformation
\begin{equation}\label{2}
\phi\rightarrow\tilde\phi=\phi+\epsilon,\qquad \kappa\rightarrow\tilde\kappa=
\kappa e^{-\alpha\epsilon}
\end{equation}
is the global isotopic symmetry of the theory with the Lagrangian (\ref{1}) for an arbitrary real value of the constant parameter $\epsilon$. We use this symmetry, which plays an important role in string theory, as the basis for further consideration. 

\section{Potential with dilaton shift symmetry
}
\mbox{}\vspace{-\baselineskip}

It is easy to see that the combination of fields 
\begin{equation}\label{2-2}
I=e^{\alpha\phi}\kappa
\end{equation}
is an invariant of transformation (\ref{2}). 
Consider now a dilaton-axion system with a Lagrangian 
\begin{equation}\label{3}
{\cal L}={\cal L}_0- {\cal V},
\end{equation} 
where ${\cal L}_0$ is given by Eq. (\ref{1}) and the potential ${\cal V}$  is defined as a function of the indicated invariant, i.e. $V=V(I)$. It is clear that the system (\ref{3}) is symmetric with respect to the dilaton shift transformation (\ref{2}). In this paper, the potential is chosen in the Higgs form:
\begin{equation}\label{4}
{\cal V}=\frac{\lambda}{4}\left(I^2-v^2\right)^2,
\end{equation}
where $\lambda$ and $v$ are positive constants.

Performing a standard analysis of this theory, we come to the conclusion that the set of vacuum states consists of pairs of constants $\{\phi_v,\,\kappa_v\}$, related to each other by the transformation (\ref{2}) and satisfying the restriction
\begin{equation}\label{5}
e^{2\alpha\phi_v}\kappa_v^2 = v^2.
\end{equation}
Indeed, it is easy to verify that the Euler-Lagrange equations corresponding to the system (\ref{1}), (\ref{2-2}), (\ref{3}) and (\ref{4}) are identically satisfied, and that the Noetherian integral of the energy of the system reaches its minimal (trivial) value on each of the representatives of this set.
	
Let us now calculate the mass spectrum of the system under consideration. To do this, we set 
\begin{equation}\label{6}
\phi=\phi_v+\xi, \qquad \kappa=\kappa_v+\eta;
\end{equation}
then for ${\cal L}_2$ -- the part of the Lagrangian quadratic in perturbation fields $\xi$ and $\eta$ over the chosen vacuum, we get: 
\begin{equation}\label{7}
{\cal L}_2= \frac{1}{2}\left[\left(\partial\xi\right)^2+e^{2\alpha\phi_v}\left(\partial\eta\right)^2 \right] - \lambda v^2 e^{2\alpha\phi_v}\left( \eta+\alpha\kappa_v\xi\right)^2.
\end{equation}
Diagonalization of Eq. (\ref{7}) is achieved by substitution
\begin{equation}\label{8}
\xi= \frac{\theta+\alpha v \chi}{\sqrt{1+\alpha^2v^2}}, \qquad \eta = e^{-\alpha\phi_v} \frac{\chi-\alpha v \theta}{\sqrt{1+\alpha^2v^2}};
\end{equation}
its result 
\begin{equation}\label{9}
{\cal L}_2= \frac{1}{2}\left(\partial\theta\right)^2 + \frac{1}{2}\left(\partial\chi\right)^2 - \frac{m_{\chi}^2}{2}\chi^2,
\end{equation} 
where 
\begin{equation}\label{10}
m_{\chi}=\sqrt{2\lambda \left(1+\alpha^2v^2\right)}\,v
\end{equation} 
shows that the system has a massless Nambu-Goldstone mode $\theta$ and a Higgs field $\chi$ with mass $m_{\chi}$ represented in Eq. (\ref{10}).

\section{Stationary case. Integrals of motion
}
\mbox{}\vspace{-\baselineskip}

Consider now the system (\ref{1}), (\ref{2-2}), (\ref{3}) and (\ref{4}) in the stationary case. We are interested in solitons, i.e. solutions that have a finite value of the energy integral \cite{R}. According to Derrick's theorem \cite{D}, there is no prohibition on their existence only for $d=1$, that is, for the original $(1+1)$-dimensional theory, which will be assumed below. Thus, starting from here $\phi=\phi(x)$ and $\kappa=\kappa(x)$, where $x=x^1$.

The equations of motion then have the following form:
\begin{equation}\label{11}
\phi''-\alpha e^{2\alpha\phi}\left[\left(\kappa'\right)^2+\lambda\left(e^{2\alpha\phi}\kappa^2-v^2\right)\kappa^2\right]=0,
\end{equation}
\begin{equation}\label{12}
\left(e^{2\alpha\phi}\kappa'\right)'-\lambda e^{2\alpha\phi}\left(e^{2\alpha\phi}\kappa^2-v^2\right)\kappa=0,
\end{equation}
where the prime means taking the derivative with respect to $x$. It is easy to check that the equations (\ref{11}), (\ref{12}) are the Euler-Lagrange equations for the Noetherian energy integral of the system (\ref{1}), (\ref{2-2}), (\ref{3}) and (\ref{4}) for $d=1$ in the stationary case. 
This integral is defined as
\begin{equation}\label{13}
{\cal E}=\int_{-\infty}^{+\infty}dx \,T_0^0,
\end{equation}
where
\begin{equation}\label{14}
T_0^0 = \frac{1}{2}\left[\left(\phi'\right)^2+e^{2\alpha\phi} \left(\kappa'\right)^2\right]+ \frac{\lambda}{4}\left(e^{2\alpha\phi}\kappa^2-v^2\right)^2
\end{equation}
is the corresponding component of the energy-momentum tensor.
	
The energy functional (\ref{13}) admits a simple Noetherian analysis, which makes it possible to integrate the system of nonlinear equations (\ref{11}), (\ref{12}). Namely, it has explicit symmetry with respect to the shift
\begin{equation}\nonumber
x\rightarrow \tilde x=x+\varepsilon
\end{equation}
for $\varepsilon ={\rm const}$ and invariant fields $\phi$ and $\kappa$. The corresponding Noetherian charge, i.e. the integral of motion of the system (\ref{11}), (\ref{12}), is the formal “energy”\, for the “mechanical system” with generalized coordinates $\phi$ and $\kappa$ in the formal “time”\, $x$:
\begin{equation}\label{15}
{\cal F} = \frac{1}{2}\left[\left(\phi'\right)^2+e^{2\alpha\phi} \left(\kappa'\right)^2\right] - \frac{\lambda}{4}\left(e^{2\alpha\phi}\kappa^2-v^2\right)^2
\end{equation}
In this case, the integral of motion (\ref{15}) is the result of the Legendre transformation from the formal “Lagrangian”\, $T^0_0$ to the formal “Hamiltonian”\, ${\cal F}$ according to the standard formula 
\begin{equation}\label{16}
{\cal F} = \frac{\partial T^0_0}{\partial \phi'}\phi'+\frac{\partial T^0_0}{\partial \kappa'}\kappa' - T^0_0.
\end{equation}
Then, the invariance of the functional (\ref{13}), (\ref{14}) with respect to the dilaton shift transformation (\ref{2}) gives one more integral of the system motion. Namely, applying the Noetherian formalism, we arrive at the integral of motion
\begin{equation}\label{17}
{\cal G} = \frac{\partial T^0_0}{\partial \phi'}\left.\frac{\partial \tilde\phi}{\partial\epsilon}\right |_{\epsilon=0}+\frac{\partial T^0_0}{\partial \kappa'}\left.\frac{\partial \tilde\kappa}{\partial\epsilon}\right |_{\epsilon=0}
\end{equation}
which after taking into account Eqs. (\ref{2}) and (\ref{14}) turns out to be equal to
\begin{equation}\label{18}
{\cal G} = \phi' - \alpha e^{2\alpha\phi}\kappa\kappa'.
\end{equation}
By a direct check, one can make sure that ${\cal F}'={\cal G}'=0$ when Eqs. (\ref{11}), (\ref{12}) are satisfied, as it should be for integrals of motion.

\section{Soliton solution: axion kink (antikink) 
}
\mbox{}\vspace{-\baselineskip}

Using the requirements for the convergence of the energy integral (\ref{13}), (\ref{14}), we obtain the following conditions on the asymptotics of the fields:
$\lim_{x\rightarrow \pm\infty}\phi(x)=\phi_{\pm\infty}, \qquad \lim_{x\rightarrow \pm\infty}\kappa(x)=\kappa_{\pm \infty},$
where the (constant) limit values of $\phi_{\pm\infty}$ and $\kappa_{\pm\infty}$ are bounded by a relation of the form (\ref{5}):
\begin{equation}\label{20}
e^{2\alpha\phi_{\pm\infty}}\kappa_{\pm\infty}^2 = v^2.
\end{equation}
Substituting 
these limit values into (\ref{15}) and (\ref{18}) 
shows that the integrals of motion corresponding to solitons are trivial:
\begin{equation}\label{21}
  {\cal F}={\cal G}=0.
\end{equation}
This fact allows us to integrate the system explicitly by solving the system of first-order ordinary differential equations (\ref{15}), (\ref{18}), (\ref{21}) with the boundary condition, for example, for $x\rightarrow +\infty$:
\begin{equation}\label{22}
\lim_{x\rightarrow +\infty}\phi(x)=\phi_0, \qquad \lim_{x\rightarrow +\infty}\kappa(x)=\kappa_0.
\end{equation}
Here, for compactness, we put $\phi_{+\infty}=\phi_0$, $\kappa_{+\infty}=\kappa_0$; of course, the introduced quantities satisfy the condition
\begin{equation}\label{23}
e^{2\alpha\phi_0}\kappa_0^2 = v^2.
\end{equation}
Namely, the equation ${\cal G}=0$ is easily integrated and leads, when (\ref{22}) is taken into account, to the following result: 
\begin{equation}\label{24}
e^{-2\alpha\phi}+\alpha^2\kappa^2= {\rm const} = e^{-2\alpha\phi_0}+\alpha^2\kappa_0^2.
\end{equation}
Then, using the relations ${\cal G}=0$, (\ref{23}) and (\ref{24}), we reduce the equation ${\cal F}=0$ to the following form:
\begin{equation}\label{25}
\left(\kappa'\right)^2=\frac{\lambda v^2 \left(1+\alpha^2v^2\right)}{2\kappa^2_0} \left(\kappa^2_0-\kappa^2\right)^2.
\end{equation}
It can be seen that Eq. (\ref{25}) is a relation that determines the axion kink (antikink) type solution
\begin{equation}\label{26}
\kappa=\kappa_0\tanh\left(\frac{x-x_0}{r_0}\right)
\end{equation}
in the case with $\kappa^2_0-\kappa^2>0$ centered at $x_0$ and with characteristic size (radius)
\begin{equation}\label{27}
r_0=\sqrt{\frac{2}{\lambda v^2 \left(1+\alpha^2v^2\right)}}.
\end{equation}
Here, in the case of $\kappa_0>0$, a kink is obtained, but if $\kappa_0<0$, it is an antikink. Obviously, in both cases $\kappa_{\pm\infty}=\pm\kappa_0$, from which, taking into account (\ref{20}), we immediately obtain that $\phi_{\pm\infty}=\phi_0$. The calculation of the dilaton field using (\ref{24}) and (\ref{26}) leads, when (\ref{23}) is taken into account, to the following result:
\begin{equation}\label{28}
\phi=\phi_0-\frac{1}{2\alpha}\log\left[1+\frac{\alpha^2v^2}{\cosh^2\left(\frac{x-x_0}{r_0}\right)}\right].
\end{equation}
	
Finally, the energy integral ${\cal E}={\cal E}(\alpha)$ for the constructed solution—the mass of the kink (antikink)—is indeed finite. Namely, substitution of the obtained solution into the defining relations (\ref{13}) and (\ref{14}) gives:
\begin{equation}\label{29}
{\cal E}(\alpha)=\Pi(\alpha){\cal E}(0),
\end{equation}
where
\begin{equation}\label{30}
\Pi(\alpha)=\frac{3}{4\alpha^3v^3}\left[\left(1+2\alpha^2v^2\right)\log\left(\sqrt{1+\alpha^2v^2}+\alpha v\right) - \alpha v\sqrt{1+\alpha^2v^2}\right],
\end{equation}
and
\begin{equation}\label{31}
{\cal E}(0)=\frac{4}{3}\sqrt{\frac{\lambda}{2}}\,v^3.
\end{equation}
Eqs. (\ref{29})--(\ref{31}) show how the dilaton-axion interaction “forms”\, the mass-energy of the found soliton. Indeed, ${\cal E}(0)$ is the mass of the kink (antikink) in the absence of this interaction, that is, at $\alpha=0$. In this case, the constructed solution describes a constant dilaton $\phi=\phi_0$, which does not contribute to the soliton mass, and a standard kink/antikink for an axion with $\kappa_0=\pm v$ (in this case, $\phi_0$ is a free parameter). In the presence of an interaction, the mass is “renormalized”,\, turning into ${\cal E}(\alpha)$, and the “renormalization coefficient”\, $\Pi(\alpha)$ does not depend on the asymptotics and has an unexpectedly nontrivial form.

\section*{Conclusion}
\mbox{}\vspace{-\baselineskip}

The dilaton-axion system with a Higgs-type potential with dilaton shift symmetry is interesting because of its closeness to applications of superstring theory. It has soliton solutions, which are generalizations of the standard kink and antikink in the standard scalar field theory with the Higgs potential. These solutions can form the basis for the study of dilaton-axion worlds on a brane \cite{R}, the decay of the false dilaton-axion vacuum \cite{vacdec}, and a number of other studies in the corresponding areas of classical and quantum field theory. 

The methods developed in this article can be applied to construct soliton solutions in dilaton-Maxwell electrodynamics (see \cite{KM}) with an appropriately selected potential. 
Interesting results related to the construction and study of kink-type solitons in other theories can be found, for example, in \cite{kink-1}, \cite{kink-2}.



\begin{thebibliography}{30}



\bibitem{Pol}
\textit{Polchinski  J.}  // String Theory. Cambridge University Press, 2005. 

\bibitem{StringsM}
\textit{Becker K., Becker M., Schwarz J. H.}  //String Theory and M-Theory. A Modern Introduction. Cambridge University Press, 2007. 

\bibitem{MS} {\it Maharana J., Schwarz J. H. } // Nucl.
Phys. В. 1993. {\bf 390}, P. 3.

\bibitem{HK-2} {\it Herrera-Aguilar A.,  Kechkin O. V.  } // Phys.
Rev. D. 1999. {\bf 59}, P. 124006.

\bibitem{K} {\it Kechkin O. V. } // Phys.
Rev. D. 2002. {\bf 65}, P. 066006.

\bibitem{D}
\textit{Derrick G. H.}  // J. Math. Phys.
1964. \textbf{5},  P. 1252.

\bibitem{R}
\textit{Rubakov V.}  // Classical Theory of Gauge Fields. Princeton University Press, 2002. 


\bibitem{vacdec}
\textit{Coleman S.}  // Phys.
Rev. D. 1977. {\bf 15}, P. 2929.

\bibitem{KM}
\textit{Kechkin O. V., Mosharev P. A.}  //  Int. J. Mod.
Phys. A. 2016. {\bf 31}, P. 1650169.

\bibitem{kink-1}
\textit{Blinov P.A., Gani T.V., Malnev A.A., Gani V.A., Sherstyukov V.B.}  //  Chaos, Solitons and Fractals. 2022. {\bf 165}, P. 112805.

\bibitem{kink-2}
\textit{Khare A., Duzgun A., Saxena A.}  //  Int. J. Mod. Phys. B. 2021 {\bf 355}, P. 2150324 .





\end{thebibliography}
\end {document}